\documentclass[12pt,draftclsnofoot,onecolumn]{IEEEtran}

\makeatletter
\newcommand{\vast}{\bBigg@{4}}
\newcommand{\vastt}{\bBigg@{6}}
\makeatother
\usepackage{color}
\usepackage{multirow}
\usepackage{graphicx}
\usepackage{epstopdf}
\usepackage[cmex10]{amsmath}
\usepackage{amssymb}

\usepackage{multirow}
\usepackage{algpseudocode}
\usepackage{amsthm}
\usepackage{cite}
\usepackage{bbm}
\theoremstyle{remark}

\makeatletter
\g@addto@macro\th@remark{\thm@headpunct{\normalfont:}}
\makeatother
\makeatletter
\newcommand{\distas}[1]{\mathbin{\overset{#1}{\kern\z@\sim}}}%
\newsavebox{\mybox}\newsavebox{\mysim}
\newcommand{\distras}[1]{%
  \savebox{\mybox}{\hbox{\kern3pt$\scriptstyle#1$\kern3pt}}%
  \savebox{\mysim}{\hbox{$\sim$}}%
  \mathbin{\overset{#1}{\kern\z@\resizebox{\wd\mybox}{\ht\mysim}{$\sim$}}}%
	}

\allowdisplaybreaks

\begin{document}

\title{EGC Reception for FSO Systems Under Mixture-Gamma Fading Channels and Pointing Errors}

\author{Nikolaos~I.~Miridakis and Theodoros~A.~Tsiftsis,~\IEEEmembership{Senior Member,~IEEE}%\\and~Corbett~Rowell,~\IEEEmembership{Senior Member,~IEEE}
%\thanks{Copyright \copyright\: 2015 IEEE. Personal use of this material is permitted. However, permission to use this material for any other purposes must be obtained from the IEEE by sending a request to pubs-permissions@ieee.org.}
%\thanks{This work was supported under a Social Policy Grant of the Nazarbayev University, Astana, Kazakhstan }
\thanks{N. I. Miridakis is with the Department of Computer Engineering, Piraeus University of Applied Sciences, 12244 Aegaleo, Greece (e-mail: nikozm@unipi.gr).}
\thanks{T. A. Tsiftsis is with the School of Engineering, Nazarbayev University, Astana 010000, Kazakhstan (e-mail: theodoros.tsiftsis@nu.edu.kz).}
%\thanks{C. Rowell is with the Dept. of Electronic \& Computer Engineering, Hong Kong University of Science \& Technology, Hong Kong and with Rohde \& Schwarz, Munich, Germany (email: corbett.rowell@gmail.com).}
}

\markboth{}%
{}

\maketitle

\begin{abstract}
The performance of equal gain combining reception at free-space optical communication systems is analytically studied and evaluated. We consider the case when the total received signal undergoes independent and not necessarily identically distributed channel fading, modeled by the versatile mixture-Gamma distribution. Also, the misalignment-induced fading due to the presence of pointing errors is jointly considered in the enclosed analysis. New closed-form expressions in terms of finite sum series of the Meijer's $G$-function are derived regarding some key performance metrics of the considered system; namely, the scintillation index, outage probability, and average bit-error rate. Based on these results, some useful outcomes are manifested, such as the system diversity order, the crucial role of diversity branches, and the impact of composite channel fading/pointing error effect onto the system performance. 
\end{abstract}


\IEEEpeerreviewmaketitle

\section{Introduction}
\IEEEPARstart{O}{ne} of the main challenges in free-space optical (FSO) communications is the channel fading due to turbulence-induced scintillation and misalignment. The former is mainly caused by the random fluctuation of the refractive index, whereas the latter by time-varying pointing errors due to thermal expansion, dynamic wind loads, and internal vibrations \cite{j:FaridHranilovic2012}. Spatial diversity can be used as a means to effectively tackle this issue; e.g., by using multiple apertures at the transceiver. Some of the most popular spatial diversity schemes are the maximum ratio combining (MRC), equal gain combining (EGC), and selection combining (SC). MRC outperforms the other schemes at the cost of an increased computational complexity, which gets more emphatic in very high data-rate communications such as FSO systems. On the other hand, SC requires less computational burden at the cost of the poorest performance. Between these two extremes, EGC presents an efficient tradeoff between performance and complexity, which is suitable for practical applications.

To cope with the composite turbulence-induced and misalignment fading, several distribution models have been proposed. Among them, the most popular ones (due to their accuracy, versatility, generality, and mathematical tractability) are the Gamma-Gamma and M\'alaga distribution models. These models, combined with the pointing error effect, result to highly complicated probability density function (PDF)/cumulative distribution function (CDF) because the rather cumbersome high-order Meijer's $G$-function is included \cite{j:AnsariYilmaz2016}. Mostly due to this reason, the performance of diversity receivers has been only merely studied so far in FSO communication systems. Specifically, the performance of dual-branch diversity receivers was studied in \cite{j:AlQuwaieeAlouini2016}, using a generalized pointing error model. In \cite{j:FaridHranilovic2012}, the performance of diversity receivers with arbitrary branches was analyzed under log-normally atmospheric fading (which efficiently approaches weak-only turbulence conditions where large-scale fluctuations dominate). However, the aforementioned works provided asymptotic results, tightly approximating the exact performance only in the high signal-to-noise (SNR) regions (i.e., $\geq 25$dB). In addition, the average bit-error rate (ABER) of EGC under the composite Gamma-Gamma fading with pointing errors was studied in \cite{j:BhatnagarGhassemlooy2016}. Yet, although the analysis in the latter work was considered for the entire SNR region of operation, it was in the form of a double infinite series representation.

In this paper, we study the performance of EGC reception for FSO systems with arbitrary diversity branches over composite fading channels and pointing errors. The mixture Gamma (MG) distribution model is used to capture the statistical properties of turbulence-induced channel fading. It was recently indicated in \cite{j:SandalidisMG2016} that MG distribution sharply coincides to both the Gamma-Gamma and M\'alaga generalized distribution models. Thus, it serves as an effective model to accurately approach channel fading in weak-to-strong turbulence conditions. Further, the misalignment-induced fading is also considered, by assuming the commonly adopted model of zero boresight pointing errors. New closed-form expressions are derived for key system performance metrics, i.e., the scintillation index, outage probability, and ABER. Moreover, some useful engineering insights are revealed, such as the system diversity order, the crucial role of diversity branches, and the impact of composite channel fading/pointing error effect onto the system performance.

\emph{Notation}: $f_{X}(\cdot)$ and $F_{X}(\cdot)$ represent PDF and CDF of the random variable (RV) $X$, respectively. Moreover, $\Gamma(\cdot)$ denotes the Gamma function \cite[Eq. (8.310.1)]{tables}, $B(\cdot,\cdot)$ is the Beta function \cite[Eq. (8.384.1)]{tables}, $\Gamma(\cdot,\cdot)$ is the upper incomplete Gamma function \cite[Eq. (8.350.2)]{tables}, and ${\rm erf(\cdot)}$ represents the error function \cite[Eq. (8.253.1)]{tables}. Finally, $G[\cdot]$ represents the Meijer's $G$-function \cite[Eq. (9.301)]{tables}.

\section{System Model}
\label{System Model}
Consider an FSO system with a single transmitter/aperture and $N$ apertures/diversity branches at the receiver. Hence, the received signal is modeled as $r=z s+\nu$, where $s \in \{0,1\}$ denotes the transmitted symbol and $\nu$ is the additive white Gaussian noise at the receiver, modeled as a complex-Gaussian RV with zero-mean and unit-variance (i.e., assume a normalized received power so as to reflect the corresponding SNR). Also, it holds that $z\triangleq \sum^{N}_{j=1}x_{j}y_{j}$, standing for the joint channel fading (irradiance) and misalignment due to the possible pointing errors at the receiver. Specifically, $x_{j}$ and $y_{j}$ are the MG channel fading and pointing error effect, respectively, at the $j^{\rm th}$ receive aperture. Moreover, assume that the RV sets $\{x_{j}\}^{N}_{j=1}$ and $\{y_{j}\}^{N}_{j=1}$ are independent and not necessarily identically distributed (i.n.n.i.d.) and independent and identically distributed (i.i.d.), correspondingly. Also, $x_{j}$ is independent of $y_{j}$ $\forall j$. The corresponding PDFs are given by \cite{j:SandalidisMG2016}
\begin{align}
f_{x_{j}}(x)=\sum^{L}_{i=1}a_{i}x^{b_{i}-1}\exp\left(-c_{i}x\right), \ x>0,
\label{fx}
\end{align}
and $f_{y_{j}}(y)=\frac{\xi^{2}}{A^{\xi^{2}}_{0}}y^{\xi^{2}-1}, \ 0\leq y\leq A_{0}$, where $a_{i}$, $b_{i}$ and $c_{i}$ denote the parameters of MG distribution. As an illustrative example, to model Gamma-Gamma channel fading, these parameters are defined as \cite[Eq. (5)]{j:SandalidisMG2016}
\begin{align}
b_{i}=\alpha,\ \ a_{i}=\frac{\frac{\alpha\beta^{\alpha}w_{i}t^{-\alpha+\beta-1}_{i}}{\Gamma(\alpha)\Gamma(\beta)}}{\left(\sum^{N}_{j=1}\frac{w_{j}t^{\beta-1}_{j}}{\Gamma(\beta)}\right)},\ \ c_{i}=\frac{\alpha\beta}{t_{i}},
\label{GGparams}
\end{align}
where $\alpha$ and $\beta$ are the small- and large-scale fading parameters, respectively. Moreover, $w_{i}$ and $t_{i}$ stand for the weight factors (ensuring that $\sum^{L}_{i=1}w_{i}=1$) and abscissas of the MG distribution, respectively \cite[p. 81]{b:NIST}. A similar matching can be easily obtained when considering another generalized fading model, namely, the M\'alaga distribution \cite[Eq. (9)]{j:SandalidisMG2016}. The total number of sum terms in \eqref{fx} determines the accuracy level of MG distribution. It was shown in \cite[Figs. 2 and 3]{j:SandalidisMG2016} that a condition of $L\leq 10$ satisfies an acceptable accuracy level for most practical applications, resulting to a rapidly converging series of \eqref{fx}. Further, $A_{0}$ is a constant term that defines the pointing loss given by $A_{0}=[{\rm erf}(\sqrt{\pi}r/(\sqrt{2}w_{z}))]^{2}$, while $r$ is the radius of the detection aperture, and $w_{z}$ is the beam waist. In addition, $\xi$ denotes the ratio between the equivalent beam radius at the receiver and the pointing error displacement standard deviation (jitter) at the receiver (i.e., the condition when $\xi\rightarrow +\infty$ reflects the non-pointing error case).

Upon the signal detection, the receiver applies EGC to enhance the communication quality, such that the corresponding SNR reads as $\gamma\triangleq \bar{\gamma}z^{2}=\bar{\gamma}(\sum^{N}_{j=1}x_{j}y_{j})^{2}$, where $\bar{\gamma}\triangleq E_{s}\bar{I}^{2}/(N N_{0})$ is the (normalized) mean SNR with $E_{s}$, $N_{0}$, and $\bar{I}$ denoting the transmit power, noise variance, and received mean irradiance, respectively.

The considered system model and corresponding analysis hereinafter can be easily extended to the case when $M\geq 1$ transmit apertures are applied, by simply substituting $N$ with $M N$ and setting $\bar{\gamma}=E_{s}\bar{I}^{2}/(M^{2}N N_{0})$ \cite[Eq. (3)]{j:BhatnagarGhassemlooy2016}.

\section{Statistical Analysis}
\label{Statistical Analysis}
We commence by extracting some key statistical results regarding the received SNR, namely, its PDF, CDF and statistical moments. These results are quite useful for the overall performance evaluation of the considered system.

The PDF of SNR for EGC-enabled reception is obtained as
\begin{align}
\nonumber
&f_{\gamma}(z)=\sum^{L}_{i_{1}=1}\cdots\sum^{L}_{i_{N}=1}\prod^{N}_{s=1}\left[a_{i_{s}}\right] \prod^{N-1}_{j=1}\left[B\left(\sum^{j}_{l=1}b_{i_{l}},b_{i_{j+1}}\right)\right]\\
&\times \frac{\xi^{2 N}c^{\xi^{2}-b_{i_{1}}}_{i_{1}}\Psi[A_{0}]z^{\frac{\sum^{N}_{l=2}b_{i_{l}}+\xi^{2}}{2}-1}}{2 A^{N \xi^{2}}_{0}\bar{\gamma}^{\sum^{N}_{l=2}b_{i_{l}}+\xi^{2}}}\Gamma\left(b_{i_{1}}-\xi^{2},\frac{c_{i_{1}}\sqrt{z}}{\bar{\gamma}A_{0}}\right),
\label{snrpdf}
\end{align} 
where $\Psi[A_{0}]\triangleq \prod^{N}_{j=2}[\frac{A^{\xi^{2}-b_{i_{j}}}_{0}}{(\xi^{2}-b_{i_{j}})}]$ for $N\geq 2$, while $\Psi[A_{0}]\triangleq 1$, for $N=1$. Also, $\xi^{2}>b_{i_{j}}$,  $2\leq j\leq N$ should hold in \eqref{snrpdf} to be a valid PDF. The proof is relegated in the Appendix.

Notice that the restriction $\xi^{2}>b_{i_{j}}$ reflects that $\xi^{2}>\alpha$, according to \eqref{GGparams}. This implies the fact that the misalignment-induced parameter due to the pointing error effect should be greater than the corresponding turbulence-induced fading parameter. However, the latter inequality is a reasonable assumption for most practical applications, since $\xi^{2}>\alpha$ holds for typical building sway-based pointing errors \cite{j:YangTsiftsisFSO2014,j:GappmairIET2011}.

The corresponding CDF of $\gamma$ reads as $F_{\gamma}(z)=\int^{z}_{0}f_{\gamma}(z)dz$. By using \eqref{snrpdf}, transforming the upper incomplete Gamma function in terms of the Meijer's G-function \cite[Eq. (8.4.16.2)]{b:PrudnikovVol3}, while utilizing \cite[Eq. (2.24.2.2)]{b:PrudnikovVol3}, we have that
\begin{align}
\nonumber
&F_{\gamma}(z)=\sum^{L}_{i_{1}=1}\cdots\sum^{L}_{i_{N}=1}\prod^{N}_{s=1}\left[a_{i_{s}}\right] \prod^{N-1}_{j=1}\left[B\left(\sum^{j}_{l=1}b_{i_{l}},b_{i_{j+1}}\right)\right]\\
\nonumber
&\times \frac{\xi^{2 N}c^{\xi^{2}-b_{i_{1}}}_{i_{1}}\Psi[A_{0}]z^{\frac{\sum^{N}_{l=2}b_{i_{l}}+\xi^{2}}{2}}}{A^{N \xi^{2}}_{0}\bar{\gamma}^{\sum^{N}_{l=2}b_{i_{l}}+\xi^{2}}}\\
&\times G^{2,1}_{2,3}\left[\frac{c_{i_{1}}\sqrt{z}}{\bar{\gamma}A_{0}}~\vline
\begin{array}{c}
\scriptstyle 1-\left(\sum^{N}_{l=2}b_{i_{l}}+\xi^{2}\right),1 \\
\scriptstyle 0,b_{i_{1}}-\xi^{2},-\left(\sum^{N}_{l=2}b_{i_{l}}+\xi^{2}\right)
\end{array}\right].
\label{Fztotalclosedform}
\end{align}

The $n^{\rm th}$ statistical moment of $\gamma$ is defined as $\mu^{(n)}_{\gamma}\triangleq \int^{\infty}_{0}x^{n}f_{\gamma}(x)dx$. Using \eqref{snrpdf} and utilizing \cite[Eq. (6.455.1)]{tables}, $\mu^{(n)}_{\gamma}$ is obtained in a closed-form expression as 
\begin{align}
\nonumber
&\mu^{(n)}_{\gamma}=\sum^{L}_{i_{1}=1}\cdots\sum^{L}_{i_{N}=1}\prod^{N}_{s=1}\left[a_{i_{s}}\right] \prod^{N-1}_{j=1}\left[B\left(\sum^{j}_{l=1}b_{i_{l}},b_{i_{j+1}}\right)\right]\\
&\times \frac{\xi^{2 N}c^{\xi^{2}-b_{i_{1}}}_{i_{1}}\Psi[A_{0}]\Gamma\left(\sum^{N}_{l=1}b_{i_{l}}+n\right)\bar{\gamma}^{-\left(\sum^{N}_{l=2}b_{i_{l}}+\xi^{2}\right)}}{A^{N \xi^{2}}_{0}\left(\sum^{N}_{l=2}b_{i_{l}}+n+\xi^{2}\right)\left(\frac{c_{i_{1}}}{\bar{\gamma}A_{0}}\right)^{\left(\sum^{N}_{l=2}b_{i_{l}}+n+\xi^{2}\right)}}.
\label{momentszclosedform}
\end{align}

\section{System Performance}
\label{System Performance}

Capitalizing on the previously derived statistics, some useful metrics that define the overall system performance are analytically evaluated.

\subsection{Scintillation Index}
The scintillation index (SI) is an important measure for evaluating the performance of FSO systems, which is defined as ${\rm SI}\triangleq \mu^{(2)}_{\gamma}/(\mu^{(1)}_{\gamma})^{2}-1$. It can be directly extracted via the closed-form moments-function provided in \eqref{momentszclosedform}.

\subsection{Outage Probability}
The outage probability is defined as the probability that the SNR falls below a certain threshold value, $\gamma_{\text{th}}$, such that $P_{\rm out}(\gamma_{\rm th})=F_{\gamma}(\gamma_{\rm th})$. With the aid of the closed-form expression of \eqref{Fztotalclosedform}, the system outage performance can be easily computed.

\subsection{Average Bit-Error Rate}
The ABER is defined as \cite[Eq. (12)]{Yilmaz2011BivariateG} $\overline{P}_{b}\triangleq \frac{Q^{P}}{2 \Gamma(P)}\int^{\infty}_{0}z^{P-1}\exp\left(-Q z\right)F_{\gamma}(z)dz$, where $P$ and $Q$ are fixed modulation-specific parameters. Inserting \eqref{Fztotalclosedform} in the latter expression, while utilizing \cite[Eq. (2.24.3.1)]{b:PrudnikovVol3}, and \cite[Eq. (9.31.1)]{tables}, ABER is presented in a closed-form expression as
\begin{align}
\nonumber
&\overline{P}_{b}=\sum^{L}_{i_{1}=1}\cdots\sum^{L}_{i_{N}=1}\prod^{N}_{s=1}\left[a_{i_{s}}\right]\frac{\xi^{2 N}\Psi[A_{0}]}{\sqrt{\pi}\Gamma(P) A^{N \xi^{2}}_{0}}\\
\nonumber
&\times \prod^{N-1}_{j=1}\left[B\left(\sum^{j}_{l=1}b_{i_{l}},b_{i_{j+1}}\right)\right]\frac{c^{\xi^{2}-b_{i_{1}}}_{i_{1}}2^{b_{i_{1}}-\xi^{2}-3}}{(\bar{\gamma}^{2}Q)^{\frac{\left(\sum^{N}_{l=2}b_{i_{l}}+\xi^{2}\right)}{2}}}\\
&\times G^{3,2}_{3,4}\left[\frac{c^{2}_{i_{1}}}{4 Q \bar{\gamma}^{2}A^{2}_{0}}~\vline
\begin{array}{c}
\scriptstyle 1-P-\frac{\left(\sum^{N}_{l=2}b_{i_{l}}+\xi^{2}\right)}{2},1-\frac{\left(\sum^{N}_{l=2}b_{i_{l}}+\xi^{2}\right)}{2},1\\
\scriptstyle 0,\frac{b_{i_{1}}-\xi^{2}}{2},\frac{b_{i_{1}}-\xi^{2}+1}{2},-\frac{\left(\sum^{N}_{l=2}b_{i_{l}}+\xi^{2}\right)}{2}
\end{array}\right].
\label{aberclosedform}
\end{align}

\subsection{Diversity Order}
In the asymptotically high SNR region, $\bar{\gamma}\rightarrow +\infty$. Hence, expanding the Meijer's $G$-function within \eqref{Fztotalclosedform} in terms of finite sum series \cite[Eq. (9.303)]{tables} and utilizing \cite[Eq. (9.100)]{tables}, the system outage probability can be tightly approximated in the high SNR regime by
\begin{align}
\nonumber
&P_{{\rm out}|\bar{\gamma}\rightarrow +\infty}(\gamma_{{\rm th}})\approx\\
\nonumber
&\sum^{L}_{i_{1}=1}\cdots\sum^{L}_{i_{N}=1}\prod^{N}_{s=1}\left[a_{i_{s}}\right] \prod^{N-1}_{j=1}\left[B\left(\sum^{j}_{l=1}b_{i_{l}},b_{i_{j+1}}\right)\right]\\
\nonumber
&\sum^{2}_{h=1}\frac{\prod^{2}_{j=1}\Gamma(\phi_{j}-\phi_{h})\Gamma(1+\phi_{h}-\zeta_{1})}{\Gamma(1+\phi_{h}-\phi_{3})\Gamma(\zeta_{2}-\phi_{h})}\\
&\times \frac{\xi^{2 N}c^{\xi^{2}-b_{i_{1}}+\phi_{h}}_{i_{1}}\Psi[A_{0}]\gamma_{\rm th}^{\frac{\sum^{N}_{l=2}b_{i_{l}}+\xi^{2}+\phi_{h}}{2}}}{A^{N \xi^{2}+\phi_{h}}_{0}\bar{\gamma}^{\sum^{N}_{l=2}b_{i_{l}}+\xi^{2}+\phi_{h}}},
\label{Poutapprox}
\end{align}
where $\boldsymbol\zeta\triangleq [1-(\sum^{N}_{l=2}b_{i_{l}}+\xi^{2}),1]$, $\boldsymbol\phi\triangleq [0,b_{i_{1}}-\xi^{2},-(\sum^{N}_{l=2}b_{i_{l}}+\xi^{2})]$, while $\zeta_{i}$ and $\phi_{i}$ denote the $i^{\rm th}$ term of $\boldsymbol\zeta$ and $\boldsymbol\phi$, correspondingly. Note that \eqref{Poutapprox} is a rather simple expression since it is given in terms of finite sum series including elementary-only functions. Thus, some useful outcomes can be emerged, such as the system diversity order. In particular, it is well-known that the outage probability in the high SNR regime approaches the form of $\bar{\gamma}^{-\mathcal{D}}$, where $\mathcal{D}$ stands for the system diversity order. Noticing from \eqref{Poutapprox} that the outage performance is dominated by $\bar{\gamma}^{-(\sum^{N}_{l=2}b_{i_{l}}+\xi^{2})+\min\{0,-(b_{i_{1}}-\xi^{2})\}}$, we have that
$P_{{\rm out}|\bar{\gamma}\rightarrow +\infty}(\gamma_{{\rm th}})\propto \bar{\gamma}^{-\left[\sum^{N}_{j=1}b_{i_{j}}\right]}$. Thereby, the system diversity order is\footnote{The second equality of \eqref{diversityorder} arises from the identity \cite[Eq. (10.30.2)]{b:NIST} and the fact that the Gamma-Gamma distribution is asymptotically affected by the minimum of its two parameters.}  
\begin{align}
\mathcal{D}=\left[\sum^{N}_{j=1}b_{i_{j}}\right]=N \times \min\{\alpha,\beta\}.
\label{diversityorder}
\end{align} 
It is noteworthy from \eqref{diversityorder} that the \emph{rate of change} of system performance is independent of the pointing error effect (in the high SNR regime), under the condition $\xi^{2}>\min\{\alpha,\beta\}$, which is in agreement with \cite{j:FaridHranilovic2012,j:AnsariYilmaz2016}. Obviously, the pointing error effect plays a crucial role to the underlying coding (array) gain regardless of the number of receive apertures.

\section{Numerical Results}
\label{Numerical Results}
In this Section, the accuracy of the proposed approach is numerically verified. For the sake of clarity and without loss of generality, the analytical results are cross-compared with numerical results modeled by simulating Gamma-Gamma faded channels. Thus, the MG parameters $a_{i}$, $b_{i}$, and $c_{i}$ are set as per \eqref{GGparams}. Also, $L=10$, $r/w_{z}=0.1$, and $[P,Q]=[0.5,1]$ (i.e., the coherent BPSK modulation scheme is considered). The considered outage threshold is $\gamma_{\rm th}=1$ (i.e., assuming a minimum target data rate of 1 bps/Hz).

\begin{figure}[!t]
\centering
\includegraphics[trim=0.2cm 0.2cm 0.2cm 0.2cm, clip=true,totalheight=0.4\textheight]{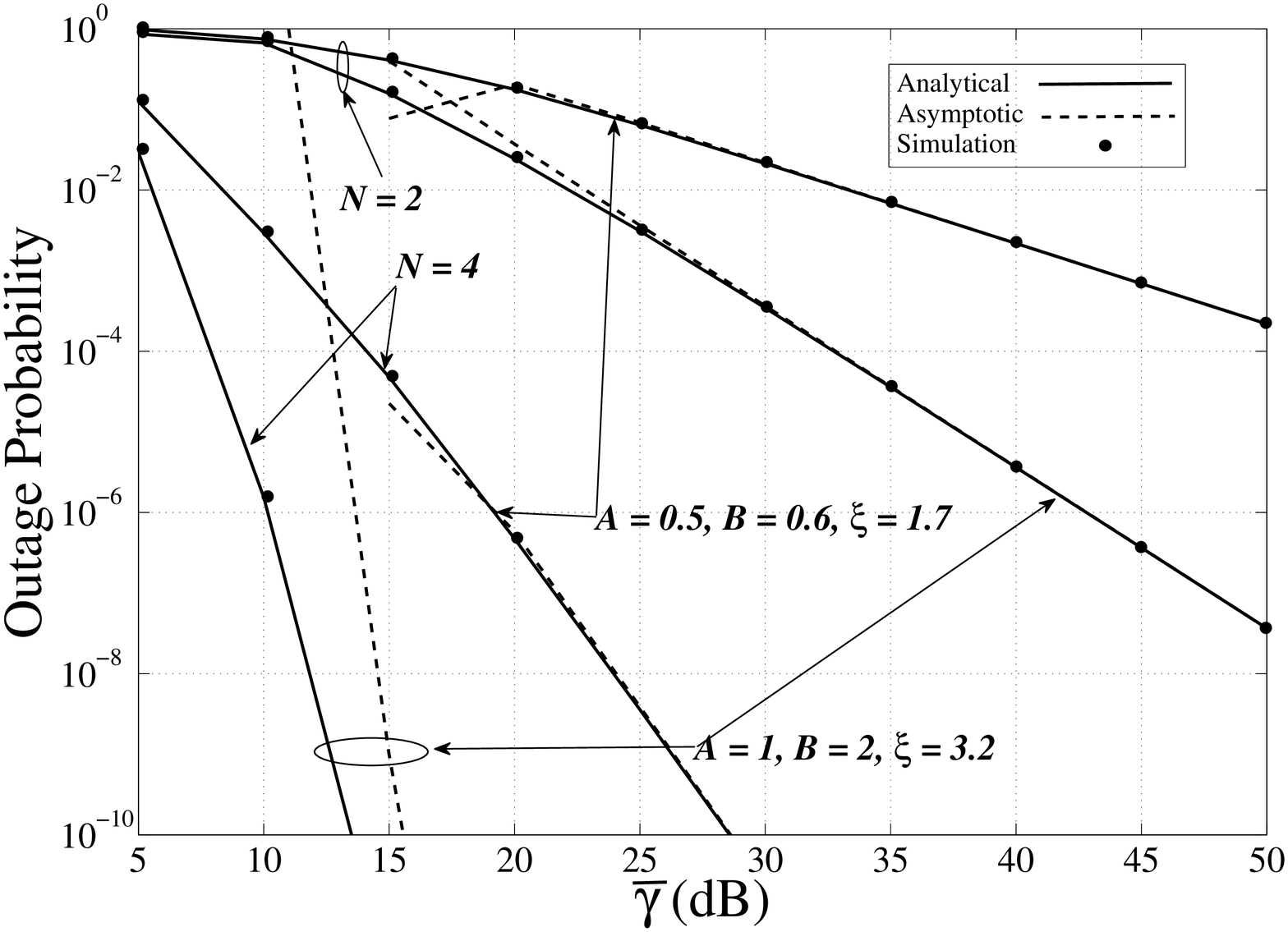}
\caption{Outage probability vs. average SNR for various system parameters.}
\label{fig1}
\end{figure}

\begin{figure}[!t]
\centering
\includegraphics[trim=0.2cm 0.2cm 0.2cm 0.2cm, clip=true,totalheight=0.4\textheight]{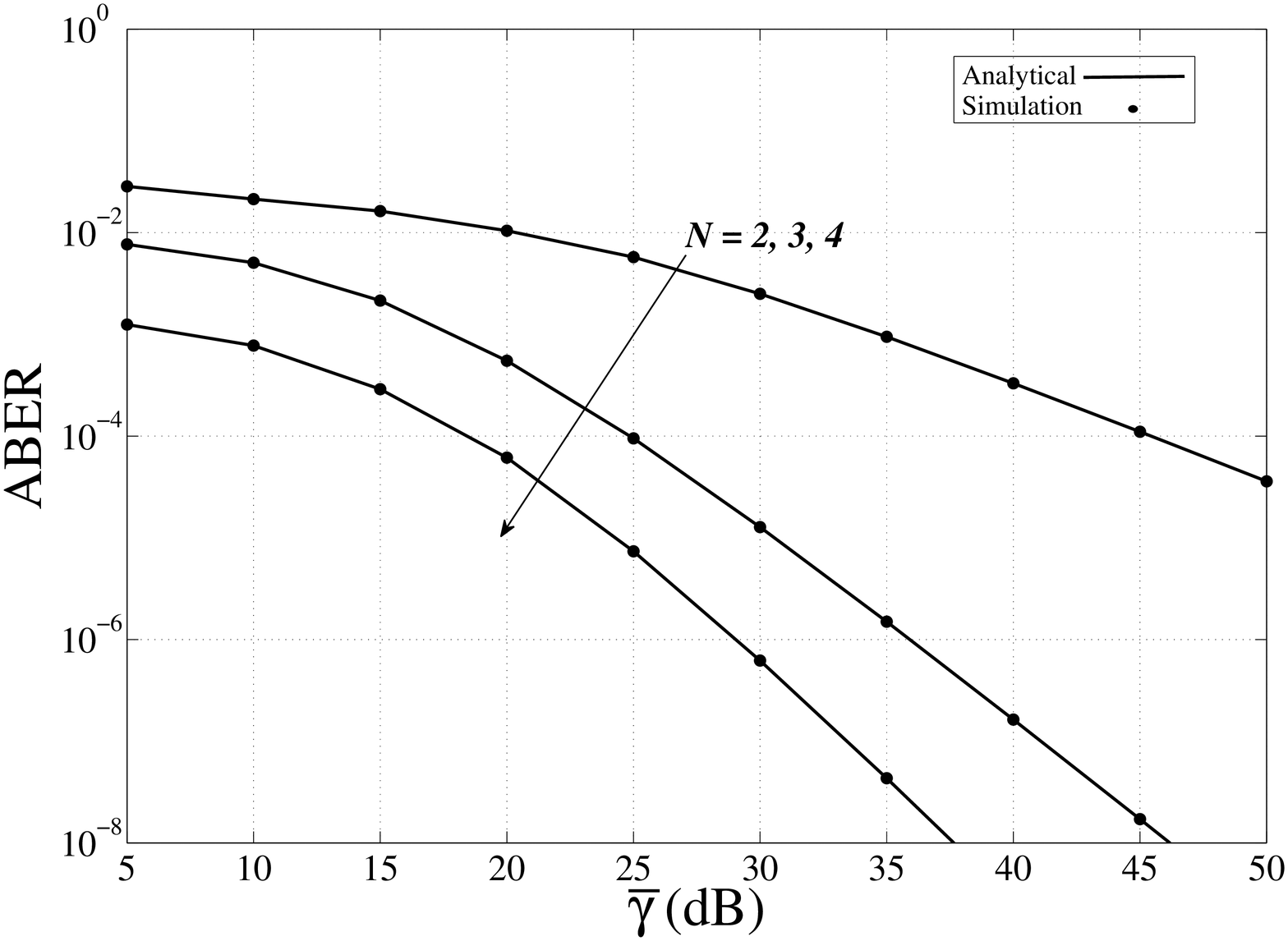}
\caption{ABER vs. average SNR for various system parameters, where $[\alpha,\beta,\xi]=[0.5,2,1]$.}
\label{fig2}
\end{figure}

The outage and ABER performance is respectively presented in Figs. \ref{fig1} and \ref{fig2} with respect to the average received SNR. Obviously, the proposed analytical approach tightly approaches the corresponding simulation numerical results at all the SNR regions and for various fading conditions. Note that lower (higher) values of the parameters $\alpha$, $\beta$, and $\xi$ indicate more (less) severe channel fading. It is also noteworthy that the asymptotic results remain efficient only in the high SNR regions (roughly, when $\bar{\gamma}>25$dB). This observation reflects on the beneficial role of the proposed scheme, which preserves the desired accuracy from low-to-high SNR values. In addition, the results of both Figs. \ref{fig1} and \ref{fig2} verify the remark of \eqref{diversityorder}. Obviously, the system performance is emphatically enhanced for a higher number of apertures, even in severe turbulence- and/or misalignment-induced fading conditions. 

\section{Conclusion}
\label{Conclusion}
The performance of EGC-enabled receivers in FSO systems was analytically studied and evaluated. Important system performance metrics were derived in closed formulation in terms of finite sum series of the Meijer's $G$-function, such as the scintillation index, outage probability, and ABER. A statistical analysis in the asymptotically high SNR regime revealed some impactful insights, such as the system diversity order, the crucial role of diversity branches, and the impact of the aforementioned fading parameters onto the overall system performance.

\appendix[Derivation of \eqref{snrpdf}]
\label{appsnr}
\numberwithin{equation}{section}
\setcounter{equation}{0}
Define $h_{j}\triangleq x_{j}y_{j}$, such that $z=\sum^{N}_{j=1}h_{j}$. Conditioning on $y_{j}$, it holds that
$f_{h_{j}|y_{j}}(x|y_{j})=\sum^{L}_{i=1}\frac{a_{i}}{y^{b_{i}}_{j}}x^{b_{i}-1}\exp(-\frac{c_{i}}{y_{j}}x)$. First, we study the case where $z=x_{1}y_{1}+x_{2}y_{2}=h_{1}+h_{2}$ and the respective PDF of $z$ is obtained as $f_{z}(z|y_{1},y_{2})=\int^{z}_{0}f_{h_{1}|y_{1}}(x|y_{1})f_{h_{2}|y_{2}}(z-x|y_{2})dx$. Utilizing \cite[Eq. (3.191.1)]{tables}, the latter integral can be evaluated as
\begin{align}
\textstyle f_{z|y_{1},y_{2}}(z|y_{1},y_{2})&=\sum^{L}_{i_{1}=1}\sum^{L}_{i_{2}=1}\frac{a_{i_{1}}a_{i_{2}}B\left(b_{i_{1}},b_{i_{2}}\right)z^{b_{i_{1}}+b_{i_{2}}-1}}{y^{b_{i_{1}}}_{1}y^{b_{i_{2}}}_{2}\exp\left(\frac{c_{i_{1}}}{y_{1}}x\right)}.
\label{fz1condyclosedform}
\end{align}
Following a similar procedure for $N$ steps, the generalized formulation for the PDF of $z$ is derived as
\begin{align}
\nonumber
&f_{z|\{y_{j}\}^{N}_{j=1}}(z|\{y_{j}\}^{N}_{j=1})=\sum^{L}_{i_{1}=1}\cdots\sum^{L}_{i_{N}=1}\prod^{N}_{s=1}\left[\frac{a_{i_{s}}}{y^{b_{i}}_{s}}\right]\\
&\times \prod^{N-1}_{j=1}\left[B\left(\sum^{j}_{l=1}b_{i_{l}},b_{i_{j+1}}\right)\right] z^{\sum^{N}_{l=1}b_{i_{l}}-1}\exp\left(-\frac{c_{i_{1}}}{y_{1}}x\right).
\label{fzcondyclosedform}
\end{align}

We proceed on the unconditional PDF of $z$. Capitalizing on the independence amongst the involved RVs, it holds that
\begin{align}
\nonumber
f_{z}(z)=\underbrace{\int^{A_{0}}_{0}\cdots\int^{A_{0}}_{0}}_{N-{\rm fold}}&f_{z|\{y_{j}\}^{N}_{j=1}}(z|\{y_{j}\}^{N}_{j=1})\\
&\times f_{y_{1}}(y_{1})\cdots f_{y_{N}}(y_{N})dy_{1}\cdots dy_{N}.
\label{fzuncondy}
\end{align}
Integrating the latter expression with respect to $y_{1}$ with the aid of \cite[Eq. 3.351.2]{tables}, we get
\begin{align}
\nonumber
&f_{z|\{y_{j}\}^{N}_{j=2}}(z|\{y_{j}\}^{N}_{j=2})=\\
\nonumber
&\sum^{L}_{i_{1}=1}\cdots\sum^{L}_{i_{N}=1}\prod^{N}_{s=1}\left[a_{i_{s}}\right] \prod^{N-1}_{j=1}\left[B\left(\sum^{j}_{l=1}b_{i_{l}},b_{i_{j+1}}\right)\right]\\
&\times \frac{\xi^{2}c^{\xi^{2}-b_{i_{1}}}_{i_{1}}}{A^{\xi^{2}}_{0}}z^{\sum^{N}_{l=2}b_{i_{l}}+\xi^{2}-1}\Gamma\left(b_{i_{1}}-\xi^{2},\frac{c_{i_{1}}z}{A_{0}}\right)\prod^{N}_{j=2}\left[y^{-b_{i_{j}}}_{j}\right].
\label{fzuncondy1closedform}
\end{align}
Next, integrating the latter PDF with respect to $y_{2}$, an integral of the following form appears as $\int^{A_{0}}_{0}y^{\xi^{2}-b_{i_{2}}-1}_{2}dy_{2}=\frac{A^{\xi^{2}-b_{i_{2}}}_{0}}{(\xi^{2}-b_{i_{2}})}$, $\xi^{2}>b_{i_{2}}$. By following the above procedure for $N-1$ steps, the total (unconditional) PDF of $z$ reads as
\begin{align}
\nonumber
&f_{z}(z)=\sum^{L}_{i_{1}=1}\cdots\sum^{L}_{i_{N}=1}\prod^{N}_{s=1}\left[a_{i_{s}}\right] \prod^{N-1}_{j=1}\left[B\left(\sum^{j}_{l=1}b_{i_{l}},b_{i_{j+1}}\right)\right]\\
&\times \frac{\xi^{2 N}c^{\xi^{2}-b_{i_{1}}}_{i_{1}}}{A^{N \xi^{2}}_{0}}\Psi[A_{0}]z^{\sum^{N}_{l=2}b_{i_{l}}+\xi^{2}-1}\Gamma\left(b_{i_{1}}-\xi^{2},\frac{c_{i_{1}}z}{A_{0}}\right),
\label{fztotalclosedform}
\end{align}
where $\Psi[A_{0}]$ is defined back in \eqref{snrpdf}. Hence, using \eqref{fztotalclosedform}, the PDF of SNR for EGC-enabled reception is provided in \eqref{snrpdf}.

\ifCLASSOPTIONcaptionsoff
  \newpage
\fi

\bibliographystyle{IEEEtran}
\bibliography{IEEEabrv,References}

\begin{thebibliography}{10}
\providecommand{\url}[1]{#1}
\csname url@samestyle\endcsname
\providecommand{\newblock}{\relax}
\providecommand{\bibinfo}[2]{#2}
\providecommand{\BIBentrySTDinterwordspacing}{\spaceskip=0pt\relax}
\providecommand{\BIBentryALTinterwordstretchfactor}{4}
\providecommand{\BIBentryALTinterwordspacing}{\spaceskip=\fontdimen2\font plus
\BIBentryALTinterwordstretchfactor\fontdimen3\font minus
  \fontdimen4\font\relax}
\providecommand{\BIBforeignlanguage}[2]{{%
\expandafter\ifx\csname l@#1\endcsname\relax
\typeout{** WARNING: IEEEtran.bst: No hyphenation pattern has been}%
\typeout{** loaded for the language `#1'. Using the pattern for}%
\typeout{** the default language instead.}%
\else
\language=\csname l@#1\endcsname
\fi
#2}}
\providecommand{\BIBdecl}{\relax}
\BIBdecl

\bibitem{j:FaridHranilovic2012}
A.~A. Farid and S.~Hranilovic, ``Diversity gain and outage probability for
  {MIMO} free-space optical links with misalignment,'' \emph{{IEEE} Trans.
  Commun.}, vol.~60, no.~2, pp. 479--487, Feb. 2012.

\bibitem{j:AnsariYilmaz2016}
I.~S. Ansari, F.~Yilmaz, and M.~S. Alouini, ``Performance analysis of
  free-space optical links over {M}\'alaga ({$M$}) turbulence channels with
  pointing errors,'' \emph{{IEEE} Trans. Wireless Commun.}, vol.~15, no.~1, pp.
  91--102, Jan. 2016.

\bibitem{j:AlQuwaieeAlouini2016}
H.~AlQuwaiee, H.~C. Yang, and M.~S. Alouini, ``On the asymptotic capacity of
  dual-aperture {FSO} systems with generalized pointing error model,''
  \emph{{IEEE} Trans. Wireless Commun.}, vol.~15, no.~9, pp. 6502--6512, Sep.
  2016.

\bibitem{j:BhatnagarGhassemlooy2016}
M.~R. Bhatnagar and Z.~Ghassemlooy, ``Performance analysis of {G}amma-{G}amma
  fading {FSO MIMO} links with pointing errors,'' \emph{J. Lightwave Technol.},
  vol.~34, no.~9, pp. 2158--2169, May 2016.

\bibitem{j:SandalidisMG2016}
H.~G. Sandalidis, N.~D. Chatzidiamantis, and G.~K. Karagiannidis, ``A tractable
  model for turbulence- and misalignment-induced fading in optical wireless
  systems,'' \emph{{IEEE} Commun. Lett.}, vol.~20, no.~9, pp. 1904--1907, Sep.
  2016.

\bibitem{tables}
I.~S. Gradshteyn and I.~M. Ryzhik, \emph{Table of Integrals, Series, and
  Products}.\hskip 1em plus 0.5em minus 0.4em\relax Academic Press, 2007.

\bibitem{b:NIST}
F.~W. Olver, D.~W. Lozier, R.~F. Boisvert, and C.~W. Clark, \emph{NIST Handbook
  of Mathematical Functions}, 1st~ed.\hskip 1em plus 0.5em minus 0.4em\relax
  New York, NY, USA: Cambridge University Press, 2010.

\bibitem{j:YangTsiftsisFSO2014}
F.~Yang, J.~Cheng, and T.~A. Tsiftsis, ``Free-space optical communication with
  nonzero boresight pointing errors,'' \emph{{IEEE} Trans. Commun.}, vol.~62,
  no.~2, pp. 713--725, Feb. 2014.

\bibitem{j:GappmairIET2011}
W.~Gappmair, ``Further results on the capacity of free-space optical channels
  in turbulent atmosphere,'' \emph{IET Communications}, vol.~5, no.~9, pp.
  1262--1267, Jun. 2011.

\bibitem{b:PrudnikovVol3}
A.~P. Prudnikov, O.~I. Marichev, and Y.~A. Brychkov, \emph{Integrals and
  Series, Vol. 3: More Special Functions}.\hskip 1em plus 0.5em minus
  0.4em\relax Gordon and Breach science publishers, 1986.

\bibitem{Yilmaz2011BivariateG}
I.~S. Ansari, S.~Al-Ahmadi, F.~Yilmaz, M.~S. Alouini, and H.~Yanikomeroglu, ``A
  new formula for the {BER} of binary modulations with dual-branch selection
  over generalized-{$K$} composite fading channels,'' \emph{{IEEE} Trans.
  Commun.}, vol.~59, no.~10, pp. 2654--2658, Oct. 2011.

\end{thebibliography}

\vfill

\end{document}